\documentclass[prc,preprint,showpacs,showkeys,nofootinbib,tightenlines]{revtex4}
\usepackage[a4paper, left=3cm,right=3cm,top=3cm,bottom=4cm]{geometry}
\usepackage{graphicx}
\usepackage{amsmath}

\newcommand{\simle}{\hspace*{0.2em}\raisebox{0.5ex}{$<$}
     \hspace{-0.8em}\raisebox{-0.3em}{$\sim$}\hspace*{0.2em}}
\newcommand{\ep}{\epsilon}

\newcommand{\g}{\gamma}

\newcommand{\si}{\sigma}
\newcommand{\la}{\lambda}
\newcommand{\simu}{\sigma^{\mu\nu}}
\newcommand{\Fmu}{F_{\mu\nu}}
\newcommand{\Gmu}{G^a_{\mu\nu}}

\newcommand{\slashT}{\slash \hspace{-0.4em}T}
\newcommand{\h}{\frac{1}{2}}

\newcommand{\qb}{\bar q}
\newcommand{\Nb}{\bar N}
\newcommand{\Fp}{F_\pi}

\newcommand{\tb}{\bar \theta}

\newcommand{\mpi}{m_{\pi}}
\newcommand{\MQ}{M_{\mathrm{QCD}}}

\newcommand{\Or}{\mathcal O}

\newcommand{\dslash}[1]{#1 \llap{/\kern-0.5pt}}
\newcommand{\Dslash}[1]{#1 \llap{/\kern+1.2pt}}
\newcommand{\DDslash}[1]{#1 \llap{/\kern+2.3pt}}
\newcommand{\dslashh}[1]{#1 \llap{/\kern+1pt}}

\newcommand{\boldtau}{\mbox{\boldmath $\tau$}}
\newcommand{\boldpi}{\mbox{\boldmath $\pi$}}

\begin{document}
\title{The Nucleon Electric Dipole Form Factor\\ 
From Dimension-Six Time-Reversal Violation}

\author{J. de Vries}
\affiliation{KVI, Theory Group, University of Groningen,
 9747 AA Groningen, The Netherlands}

\author{E. Mereghetti}
\affiliation{Department of Physics, University of Arizona,
 Tucson, AZ 85721, USA}

\author{R. G. E. Timmermans}
\affiliation{KVI, Theory Group, University of Groningen,
 9747 AA Groningen, The Netherlands}

\author{U. van Kolck}
\affiliation{Department of Physics, University of Arizona,
 Tucson, AZ 85721, USA}

\date{\today}

\begin{abstract}
We calculate the electric dipole form factor of the nucleon
that arises as a low-energy manifestation
of time-reversal violation in quark-gluon interactions
of effective dimension 6: 
the quark electric and chromoelectric dipole moments,
and the gluon chromoelectric dipole moment.
We use the framework of two-flavor
chiral perturbation theory to one loop.
\end{abstract}

\maketitle

Electric dipole moments (EDMs) \cite{KhripLam1997,Pospelov:2005pr}
provide stringent bounds on sources of time-reversal ($T$)
violation beyond the phase of the quark-mixing matrix \cite{Kobayashi:1973fv}.
Experiments are in preparation \cite{expts} which aim
to improve the current bound on the neutron
EDM, $|d_n| < 2.9 \cdot 10^{-26} \, e$ cm \cite{dnbound},
by nearly two orders of magnitude.
Novel ideas exist \cite{storageringexpts} also for the measurement of EDMs 
of charged particles in storage rings, including 
the proton ---for which an indirect bound,
$|d_p|< 7.9 \cdot 10^{-25} \, e$ cm,
has been extracted from 
the atomic Hg EDM \cite{hgbound}--- and the deuteron.
Since the Standard Model prediction \cite{McKellar:1987tf,Pospelov:1994uf} 
is orders of 
magnitude away from 
current experimental limits, a signal in this new crop of experiments
would be a clear sign of new physics. 

The momentum dependence of
an EDM is the electric dipole form factor (EDFF).
Together with the well-known 
parity ($P$) and $T$-preserving
electric and magnetic form factors and the $P$-violating, $T$-preserving
anapole form factor, the $P$- and $T$-violating EDFF completely specifies 
the Lorentz-invariant electromagnetic current of a particle with spin 1/2. 
Although the full momentum dependence of a nuclear EDFF
will not be measured anytime soon, the radius of the form factor
provides a contribution to the Schiff moment (SM)  of the corresponding
atom, because it produces a short-range electron-nucleus interaction.

There has been some recent interest \cite{BiraHockings,Faessler,Ottnad}
on the nucleon EDFF stemming
from the lowest-dimension $T$ violation in strong interactions, 
the QCD $\bar{\theta}$ term.
As other low-energy observables, both the EDM and the SM of hadrons and 
nuclei are difficult to calculate directly in QCD. 
Attempts have been made to extract the nucleon EDM from
lattice simulations \cite{lattice}, but a signal with dynamical quarks
remains elusive.
One possible way to extract the EDM in this case relies on a extrapolation
of the EDFF to zero momentum,
which provides another motivation to look at the EDFF.
QCD-inspired models have also been brought to bear
on the nucleon EDFF \cite{Faessler}.

We would like to use a framework flexible enough to formulate
the nucleon EDFF in the wider context of other low-energy $T$-violating
observables such as the EDMs of nuclei. 
Such framework exists in the form of an effective field theory,
chiral perturbation theory (ChPT) \cite{weinberg79,Jenkins:1990jv,original}.
(For introductions, see for example Refs. \cite{Weinberg,Bernard:1995dp}.)
Since it correctly incorporates 
the approximate chiral symmetry of QCD, 
ChPT provides not only a model-independent
description of low-energy physics but also the quark-mass
dependence of observables, which is useful in the extrapolation
of lattice results to realistic values of the pion mass.
The nucleon EDFF from the $\bar{\theta}$ term has in fact been
calculated in this framework \cite{BiraHockings,Ottnad}, and some
implications of the particular way the $\bar{\theta}$ term
breaks chiral symmetry 
were discussed in Ref. \cite{BiraEmanuele}.
(For earlier work on the neutron EDM in ChPT, see for example
Refs. \cite{CDVW79,su3}.)
The momentum dependence of the EDFF is given by the 
pion cloud \cite{Thomas:1994wi,BiraHockings}:
the scale for momentum variation is the pion mass and
the SM is determined
by a $T$-violating pion-nucleon coupling.
Assuming naturalness of ChPT's low-energy constants (LECs),
one can use an estimate of
this coupling 
based on $SU(3)$ symmetry to derive \cite{CDVW79}
a bound on $\bar{\theta}$, 
$\bar{\theta}\simle 2.5 \cdot 10^{-10}$ \cite{Ottnad}
from the current limit on the neutron EDM.
ChPT extrapolation formulas for the nucleon EDM in lattice QCD
can be found in Ref. \cite{oconnell}.

The smallness of $\bar{\theta}$ leaves room for other sources
of $T$ violation in the strong interactions. 
Here we calculate in ChPT the nucleon EDFF arising from the effectively 
dimension-6 interactions involving quark and gluon fields
that violate $T$ \cite{Rujula,Weinberg:1989dx}: 
the quark electric dipole moment (qEDM), which couples quarks and photons; 
the quark chromoelectric dipole moment (qCEDM), which couples quarks and 
gluons; 
and the Weinberg operator, which couples three gluons and 
can be identified as the gluon chromoelectric dipole moment (gCEDM). 
These higher-dimension
operators can have their origin in an ultraviolet-complete 
theory at a high-energy
scale, such as, for example, supersymmetric 
extensions of the Standard Model.
We construct the interactions among nucleons, pions and
photons that stem from the underlying quark-gluon operators 
and use them to calculate the EDFF to the order
where the momentum dependence first appears.
As we will see, the sizes of the proton and neutron
EDMs and SMs partially reflect the underlying sources of $T$ violation.
While much effort has already been put into estimating the EDMs from
these sources \cite{KhripLam1997,Pospelov:2005pr},
the full EDFF apparently has been previously considered only
within a particular chiral quark model \cite{otherFaessler}.
Other implications of the different chiral transformation properties \cite{will}
of the dimension-6 operators will be considered in a subsequent paper
\cite{morejordy}.

Well below the scale $M_{\slashT}$ characteristic of $T$ violation,
we expect $T$-violating effects to be captured by the
lowest-dimension interactions
among Standard Model fields that respect the theory's
$SU(3)_c\times SU(2)_L\times U(1)_Y$ gauge symmetry.
Just above the characteristic QCD scale $\MQ \sim 1$ GeV, 
strong interactions are described by the
most general Lagrangian with Lorentz, and color and electromagnetic
gauge invariance among 
the lightest quarks ($q=(u \; d)^T$), gluons ($G_{\mu}^a$), 
and photons ($A_{\mu}$).
The effectively dimension-6 $T$-violating terms at this scale can be written as
\begin{eqnarray}
{\cal L}_{\slashT}
&=&
-\frac{i}{2}\qb \left(d_0+d_3 \tau_3\right)\simu \g^5 q \; \Fmu
-\frac{i}{2}\qb \left(\tilde{d}_0+\tilde{d}_3 \tau_3\right)\simu\g^5\la^a q 
\; \Gmu
\nonumber\\
&&+\frac{d_W}{6}\ep^{\mu\nu\la\si}f^{abc}
G^a_{\mu \rho}G_\nu^{b,\rho}G^c_{\la \si},
\label{eq:dim6}
\end{eqnarray}
in terms of the photon and gluon field strengths $\Fmu$ and $\Gmu$, 
respectively,
the standard products of gamma matrices $\g^5$ and $\simu$ in spin space,
the totally antisymmetric symbol $\ep^{\mu\nu\la\si}$,
the Pauli matrix $\tau_3$ in isospin space,
the Gell-Mann matrices $\la^a$ in color space,
and the associated Gell-Mann coefficients
$f^{abc}$. 
In Eq. (\ref{eq:dim6}) the first (second) term represents
the isoscalar $d_0$ ($\tilde{d}_0$) and isovector $d_3$ ($\tilde{d}_3$)
components of the qEDM (qCEDM). Although these interactions have canonical
dimension 5, they originate just above the Standard Model scale $M_{W}$
from dimension-6 operators  \cite{Rujula} involving 
in addition the carrier of electroweak symmetry breaking 
(the Higgs field). 
They are thus proportional to the vacuum expectation value of the Higgs field,
which we can trade for the ratio 
of the quark mass to Yukawa coupling, $m_q/f_q$.
Writing the proportionality constant as 
$e\delta_q f_q/M_{\slashT}^{2}$ ($4\pi \tilde{\delta}_q f_q/M_{\slashT}^{2}$),
\begin{eqnarray}
d_i \sim \Or\left(e \delta
\frac{\bar m}{M_{\slashT}^2}\right),
\qquad
\tilde d_i\sim \Or\left(4\pi \tilde \delta 
\frac{\bar m}{M_{\slashT}^2}
\right),
\end{eqnarray}
in terms of the average 
light-quark mass $\bar{m}$ and the dimensionless factors 
$\delta$ and $\tilde{\delta}$ representing typical values 
of $\delta_q$ and $\tilde{\delta}_q$.
The third term in Eq. (\ref{eq:dim6}) 
\cite{Weinberg:1989dx} is the gCEDM, with 
\begin{eqnarray}
d_W \sim \Or\left(\frac{4\pi w}{M^2_{\slashT}}\right)
\end{eqnarray}
in terms of a dimensionless parameter $w$.
The sizes of 
$\delta$, $\tilde{\delta}$ and $w$
depend on the exact mechanisms of electroweak and $T$ breaking
and on the running to the low energies where non-perturbative QCD effects
take over.
The minimal assumption
is that they are $\Or(1)$, $\Or(g_s/4\pi)$ and $\Or((g_s/4\pi)^3)$,
respectively, with 
$g_s$ the strong-coupling constant.
However they can be 
much smaller (when parameters encoding $T$-violating beyond the Standard Model
are small) or much larger (since $f_q$ is unnaturally small).
In the Standard Model itself,
where $M_{\slashT}=M_{W}$, they are suppressed \cite{Pospelov:1994uf}
by the Jarlskog parameter \cite{Jarlskog} $J_{CP}\simeq 3 \cdot 10^{-5}$.
In supersymmetric models with various simplifying,
universality assumptions 
of a soft-breaking sector with a common scale $M_{SUSY}$, 
one has $M_{\slashT}=M_{SUSY}$ and 
the size of the dimensionless parameters is given by the minimal assumption times
a factor which is \cite{arnowitt,ibrahim,Pospelov:2005pr}, 
roughly (neglecting electroweak parameters),
$A_{CP}=(g_s/4\pi)^2 \sin \phi$,
with $\phi$ a phase encoding $T$ violation.
Allowing 
for non-diagonal terms in the soft-breaking sfermion mass matrices,
enhancements of the type $m_b/m_d\sim 10^3$ or even $m_t/m_u\sim 10^5$
are possible (although they might be associated with
other, smaller phases) \cite{Pospelov:2005pr}.

Since we are interested in light systems,
we are integrating out all degrees of freedom associated with quarks heavier
than up and down. The effects of qEDMs and qCEDMs of such quarks are
discussed briefly at the end.
$T$-violating four-quark operators are effectively dimension-8
because again 
electroweak gauge
invariance requires insertions of the Higgs field.
Since higher-dimension operators are suppressed by more inverse powers 
of the large scale $M_{\slashT}$, we expect them to be generically
less important at low energies and we 
concentrate here on the dimension-6 operators
in Eq. (\ref{eq:dim6}).
It is of course
possible that in particular models the coefficients of the 
effectively dimension-6 operators 
are suppressed enough to make 
higher-dimension operators numerically important;
low-energy implications of four-quark operators,
which also contain representations of chiral symmetry 
we consider, have recently been studied
in Ref. \cite{ji}.

At momenta $Q$ comparable to the pion mass, $Q\sim m_\pi\ll \MQ$,
interactions among nucleons, pions and photons are described
by the most general Lagrangian involving these
degrees that transforms properly under the symmetries of the QCD.
Ignoring quark masses and charges and the $\bar{\theta}$ term,
the dimension-4 QCD terms are invariant under a chiral
$SU(2)_L\times SU(2)_R\sim SO(4)$ symmetry,
which is spontaneously broken down to its diagonal, isospin subgroup
$SU(2)_{V}\sim SO(3)$.
The corresponding Goldstone bosons are identified as the pions,
which provide a non-linear realization of chiral symmetry.
Pion interactions proceed through a covariant derivative,
which in stereographic coordinates \cite{Weinberg} $\boldpi$ for the pions
is written as
\begin{equation}
D_\mu \boldpi =D^{-1} \partial_\mu \boldpi,
\end{equation}
with $D=1+ \boldpi^2/F_\pi^2$ and $F_\pi\simeq 186$ MeV the pion
decay constant.
Nucleons are described by an isospin-1/2 field $N$, and 
the nucleon covariant derivative is
\begin{equation}
{\mathcal D}_\mu N =\left(\partial_\mu 
 +\frac{i}{F_\pi^2}\boldtau\cdot \boldpi \times D_\mu\boldpi\right)N.
\end{equation}
We define ${\mathcal D}^\dagger$ through
$\bar N \mathcal D^\dagger\equiv \overline{\mathcal D N}$, and 
use the shorthand notation
\begin{equation}
\mathcal D_\pm^\mu \equiv \mathcal D^\mu \pm \mathcal D^{\dagger\mu} , 
\qquad 
\mathcal D_\pm^\mu \mathcal D_\pm^\nu \equiv
\mathcal D^\mu \mathcal D^\nu
+\mathcal D^{\dagger \mu} \mathcal D^{\dagger \nu}
\pm \mathcal D^{\dagger\mu} \mathcal D^\nu
\pm \mathcal D^{\dagger\nu} \mathcal D^\mu ,
\end{equation}
and 
\begin{equation}
\tau_i \mathcal D_\pm^\mu \equiv 
\tau_i\mathcal D^\mu \pm\mathcal D^{\dagger \mu} \tau_i,
\qquad
\tau_i \mathcal D_\pm^\mu \mathcal D_\pm^\nu\equiv 
\tau_i \mathcal D^\mu \mathcal D^\nu
+\mathcal D^{\dagger \mu} \mathcal D^{\dagger \nu}\tau_i 
\pm \mathcal D^{\dagger\mu} \tau_i \mathcal D^\nu
\pm \mathcal D^{\dagger\nu} \tau_i \mathcal D^\mu .
\end{equation}
Covariant derivatives of covariant derivatives
can be constructed similarly,
for example 
\begin{equation}
(\mathcal D_{\mu} D_{\nu} \pi)_i = \left(\partial_{\mu} \delta_{i j} 
- \frac{2}{F^2_{\pi}} \varepsilon^{i k j} (\boldpi \times D_{\mu} \boldpi)_k 
\right) D_{\nu} \pi_j.
\end{equation}
For simplicity we omit the delta isobar here,
although one can introduce  \cite{vanKolck} an isospin-3/2 field for it
along completely analogous lines.
The effective interactions are constructed as isospin-invariant
combinations of chiral-covariant objects \cite{Weinberg}.

The quark mass, charge and $\bar\theta$ terms 
break chiral symmetry explicitly as specific components of
various chiral tensors.
The formalism to include chiral-symmetry-breaking operators in the 
$SU(2)\times SU(2)$ ChPT 
Lagrangian has been developed in Refs. \cite{Weinberg,vanKolck}.
Introducing the $SO(4)$ vectors
\begin{eqnarray}
\begin{array}{lcr}
S=\left(\begin{array}{c}-i\qb \g^5 \boldtau q\\
\qb q\end{array}\right),
\qquad P=\left(\begin{array}{c}\qb \boldtau q\\
i\qb\g^5 q\end{array}\right),
\end{array}
\label{eq:vectors}
\end{eqnarray}
and the $SO(4)$ scalar and antisymmetric tensor  
\begin{equation}\label{eq:tensors}
I^{\mu}=\frac{1}{6} \qb \g^\mu q,
\qquad T^{\mu}=\h \left(\begin{array}{cc}
\ep_{ijk}\qb \g^{\mu}\g^5 \tau_k q &\qb \g^\mu \tau_j q\\
-\qb \g^\mu \tau_i q              &0
\end{array} \right),
\end{equation}
the average quark-mass term transforms as $S_4$, 
the quark-mass-difference term as $P_3$,
the quark-photon coupling as $I\oplus T_{34}$,
and the $\bar\theta$ term as $P_4$.
They generate interactions containing the pion field explicitly,
which are proportional to powers of the symmetry-breaking parameters 
$\bar{m}=(m_u+m_d)/2$, $\varepsilon \bar{m}=(m_d-m_u)/2$,
$e$ (the proton charge),
and $({\bar m} (1-\varepsilon^2)\sin \tb)/2$.
The most important chiral-breaking term is the $\bar m$ term,
which among other effective interactions generates the main contribution
to the pion mass, $m^2_{\pi} = {\mathcal O}(\bar{m}\MQ)$.
The electromagnetic coupling produces two types of effective interactions:
{\it i)} purely hadronic interactions proportional to 
$\alpha_{\rm em}/4\pi\sim \varepsilon m_\pi^3/\MQ^3$;
and {\it ii)} gauge-invariant interactions with
explicit soft photon fields, which appear either in 
gauge-covariant derivatives or through the photon field strength.
The covariant derivatives below are all to be
interpreted as gauge-covariant derivatives.
After a suitable chiral rotation
eliminates it in favor of a mass term that does not generate vacuum instability
in first order in the symmetry-breaking parameter \cite{Baluni},
the $\bar\theta$ term is found to break chiral symmetry 
as a different component of the same vector $P$ to which the isospin-breaking 
quark mass term is associated.  
The construction of the corresponding effective interactions
has been carried out in some detail recently \cite{BiraEmanuele}. 
Since effective interactions
proportional to two or more powers of $T$-violating parameters
are exceedingly small, to a very good approximation
one can simply add the contributions
from dimension-6 sources considered here to the 
corresponding $\bar\theta$ contributions calculated in Refs. 
\cite{BiraHockings,Ottnad}.

Since nucleons are essentially nonrelativistic for 
$Q\ll m_N$, the nucleon mass, we work in the 
heavy-baryon framework \cite{Jenkins:1990jv}
where, instead of gamma matrices, it is the nucleon 
velocity $v^\mu$ and spin $S_\mu$
($S=(\vec{\sigma}/2, 0)$ in the rest frame $v=(\vec{0}, 1)$)
that appear in interactions.
Below we use a subscript $\perp$ to denote the component of a four-vector 
perpendicular to the velocity, for example
\begin{equation}
\mathcal D^{\mu}_{\perp} =\mathcal D^{\mu} - v^{\mu} v \cdot \mathcal D.
\end{equation}
We use reparametrization invariance (RPI) \cite{ManoharLuke} 
to incorporate Lorentz invariance in an expansion in powers of $Q/m_N$.

The infinite number of effective Lagrangian 
terms can be grouped
into sets ${\cal L}^{(\Delta)}$
of a given ``chiral index'' \cite{original} $\Delta=d+f/2-2$,
where $d$ counts derivatives, powers of $m_\pi$ and photon fields,
and $f=0, 2$ is the number of fermion fields:
\begin{equation}
 {\cal L}=\sum_{\Delta=0}^{\infty}{\cal L}^{(\Delta)}.
\label{ChiralL}
\end{equation}
The LECs 
can be estimated using naive dimensional analysis (NDA) 
\cite{NDA,Weinberg:1989dx},
in which case the index $\Delta$ tracks the number of inverse powers
of $\MQ\sim 2\pi F_\pi\simeq 1.2$ GeV associated with an interaction.
(Note that 
since NDA associates the LECs of chiral-invariant operators to $g_s/4\pi$, 
for consistency one should take $g_s\sim 4\pi$.)
For the purposes of our calculation, we need explicitly 
only the leading $T$-conserving interactions,
\begin{equation}
{\mathcal L}^{(0)} = 
\frac{1}{2} D_{\mu} \boldpi \cdot D^{\mu} \boldpi 
-\frac{m^2_{\pi}}{2D} \boldpi^2
+ \bar{N}\left( iv\cdot {\mathcal D}
         -\frac{2 g_A}{F_\pi}S^{\mu} \boldtau\cdot D_{\mu} \boldpi\right) N,
\label{LagrCons}
\end{equation}
where $g_A\simeq 1.267$ is the pion-nucleon axial-vector coupling. 
At this order the nucleon is static; kinetic corrections have relative
size ${\cal O}(Q/\MQ)$ and appear in ${\mathcal L}^{(1)}$.
Isospin breaking from the quark masses, represented by $\varepsilon$,
also only appears in subleading orders \cite{vanKolck}.

The dimension-6 sources of $T$ violation generate further effective 
interactions, which break chiral symmetry in their own ways.
Introducing the $SO(4)$ singlet
\begin{equation}\label{IW}
I_W=\frac{1}{6}\ep^{\mu\nu\la\si}f^{abc}
G^a_{\mu \rho}G_\nu^{b,\rho}G^c_{\la \si},
\end{equation}
and the $SO(4)$ vectors
\begin{eqnarray}
\begin{array}{lcr}
W=\h\left(\begin{array}{c}-i\qb \simu \g^5 \boldtau q\\
\qb\simu q\end{array}\right)\Fmu,& &
V=\h\left(\begin{array}{c}\qb \simu \boldtau q\\
i\qb\simu \g^5 q\end{array}\right)\Fmu,
\end{array}
\end{eqnarray}
and
\begin{eqnarray}
\begin{array}{lcr}
\tilde{W}=\h\left(\begin{array}{c}-i\qb \simu \g^5\boldtau \la^a q\\
\qb\simu \la^a q\end{array}\right)\Gmu,& &
\tilde{V}=\h\left(\begin{array}{c}\qb \simu \boldtau \la^a q\\
i\qb\simu \g^5 \la^a q\end{array}\right)\Gmu,
\end{array}
\end{eqnarray}
Eq. (\ref{eq:dim6}) can be rewritten as
\begin{equation}
\mathcal{L}_{\slashT}=
-d_0 V_4 +d_3 W_3-\tilde d_0 \tilde{V}_4 +\tilde d_3 \tilde{W}_3 +d_W I_W.
\label{eq:dim6.3}
\end{equation}

The corresponding $T$-violating effective Lagrangian can be constructed by
writing down all terms that transform in the same way under Lorentz, 
$P$, $T$, and chiral symmetry as the terms in Eq. \eqref{eq:dim6.3}.
We still use NDA and label operators by a generalized chiral index $\Delta$
that continues to count inverse powers of $\MQ$.
For simplicity we will not keep track of explicit factors of $\varepsilon$.
Here we present only the interactions needed in the calculation
of the nucleon EDFF up to the order the SM first appears;
as we are going to see, this means up to $\Delta =1$ for qCEDM and gCEDM
and to $\Delta =3$ for qEDM.
In the equations below, ``\ldots'' account for interactions
not needed in our calculation;
we leave a more complete presentation of the effective Lagrangian 
for a future publication \cite{morejordy}.

Some of the contributions to the EDFF
arise from virtual pions. In the presence of $T$ violation,
pions can be annihilated into the vacuum when 
an operator with the quantum numbers of the neutral pion
is allowed by the pattern of 
symmetry breaking. 
In the case of the qCEDM, such a tadpole arises from
$\tilde{W}_3$ and can thus be linear in $\tilde{d}_3$.
In the case of the gCEDM, the tadpole arises from the tensor
product of $I_W$ with the $P_3$ in Eq. (\ref{eq:vectors})
and is linear in $\varepsilon \bar{m} d_W$.
In both cases these tadpoles first appear at
$\Delta=-2$ and exist also at higher orders.
For the qEDM, they are beyond the
order we consider here because they
are suppressed by at least one factor of $\alpha_{\rm em}/4 \pi$.
All such tadpoles represent vacuum misalignment.
Because they are small, they can be treated in perturbation theory
or simply eliminated using the chiral rotation given in 
Ref. \cite{BiraEmanuele}.
To the order we are working the effects of this rotation 
can be absorbed in terms that already exist in the effective Lagrangian.
The fields and LECs introduced below are to be interpreted
as subsequent to the rotation.

Pions contribute to the EDFF in loops, which require the $T$-violating 
pion-nucleon
interactions with $\Delta = -1$.
Again, ``indirect'' electromagnetic operators stemming from
hard photons tied to qEDM are of higher order.
From the qCEDM and the gCEDM,
\begin{equation}
{\cal L}_{\slashT, \pi N}^{(-1)}=
-\frac{\bar g_0}{\Fp D}\boldpi\cdot \Nb \boldtau N
-\frac{\bar \imath_0}{\Fp^2}(v\cdot D\boldpi \times D_\mu \boldpi) \cdot 
\Nb S^\mu \boldtau N +\ldots
\label{eq:cedm1}
\end{equation}
The non-derivative term in Eq. (\ref{eq:cedm1})
arises either directly from $\tilde{V}_4$ 
or from the tensor product
$I_W \otimes S_4$, and thus has a LEC
\begin{equation}
\bar g_0= \Or\left({\tilde \delta}\frac{\mpi^2}{M_{\slashT}^2}\MQ,
                            w\frac{\mpi^2}{M_{\slashT}^2}\MQ\right) .
\end{equation}
The two-derivative term 
is the lowest-index chiral invariants that arise from $I_W$,
its LEC being
\begin{equation}
\bar \imath_{0}=\Or\left(\frac{w}{M_{\slashT}^2}\MQ\right).
\end{equation}
Note that the qCEDM and the qEDM generate two-derivative
interactions of different form than above, since they are chiral-breaking,
but they only appear at higher order.
There are other pion-nucleon interactions with $\Delta=-1$, 
but they do not contribute to the EDFF at the order we calculate.

In addition to the long-range contributions from virtual pions,
the EDFF is sensitive to shorter-range effects, which in ChPT are represented
by contact interactions.
The lowest-order contribution of this type
arises from the gCEDM combined with the quark-photon coupling,
$I_W \otimes (I\oplus T_{34})$:
\begin{equation}
{\cal L}_{\slashT, \gamma N}^{(-1)}= 
2\Nb \left\{\bar{D}_0^{(-1)}+\bar{D}_1^{(-1)}
\left[\tau_3+\frac{2}{\Fp^2 D}
      \left(\pi_3\boldpi\cdot\boldtau-\boldpi^{\,2}\tau_3\right)
           \right]\right\} 
S^\mu  N \, v^\nu \Fmu +\ldots,
\label{Wphot}
\end{equation}
where the LECs are
\begin{eqnarray}
\bar{D}_i^{(-1)}=\Or\left(\frac{ew}{ M_{\slashT}^2 }\MQ \right) .
\end{eqnarray}
In next order, there is a recoil correction
\begin{eqnarray}
{\cal L}_{\slashT, \gamma N}^{(0)}&=&
\frac{i}{m_N}
\Nb \left(\bar{D}_0^{(-1)}+\bar{D}_1^{(-1)}\tau_3\right)
S^\mu  \mathcal D^\nu_{\perp -} N \, \Fmu 
+\ldots,
\label{Wphot2}
\end{eqnarray}
and one further order up other sources contribute as well:
\begin{eqnarray}
{\cal L}_{\slashT, \gamma N}^{(1)}&=&
2\Nb\left[\bar{D}_0^{(1)}\left(1-\frac{2\boldpi^2}{F_\pi^2 D} \right)
+\bar{D}_1^{(1)}
 \left(\tau_3-\frac{2\pi_3}{F_\pi^2 D}\boldpi\cdot\boldtau\right)
\right] S^\mu N \, v^\nu \Fmu
\nonumber\\
&&+ 2\bar{D}_1^{(1)'} \left(1-\frac{2\boldpi^2}{F_\pi^2D}\right)
\Nb\left[\tau_3
         +\frac{2}{F_\pi^2D}\left(\pi_3\boldpi\cdot\boldtau-\boldpi^2\tau_3
         \right)\right]
S^\mu N \, v^\nu \Fmu
\nonumber\\
&&
-
\bar N \left({\bar S}'^{(1)}_{0}+{\bar S}'^{(1)}_{1}\tau_3\right)
\left(S \cdot \mathcal D_{\perp +} \mathcal D_{\perp +}^{\mu}
      +S^\mu \mathcal D_{\perp +}^2\right) N 
\, v^\nu \Fmu
\nonumber\\
&& 
-\frac{1}{4m_N^2}  
\Nb  \left(\bar{D}_0^{(-1)}+\bar{D}_1^{(-1)}\tau_3\right)
S \cdot \mathcal D_{\perp -}
\mathcal D_{\perp -}^{\mu}N \,  v^{\nu} \Fmu 
+\ldots
\label{eq:cedmEMI}
\end{eqnarray}
Here the operators with LECs 
\begin{eqnarray}
\bar{D}_i^{(1)}&=&
  \Or\left(e{\tilde \delta}\frac{\mpi^2}{M_{\slashT}^2} \MQ^{-1},
           e\delta\frac{\mpi^2}{M_{\slashT}^2}\MQ^{-1},
           ew \frac{\mpi^2}{M_{\slashT}^2}\MQ^{-1}\right)
\label{photonscalingC1}
\end{eqnarray}
transform as vectors:
the isoscalar component as $V_4$
or as the vectors in $\tilde{V}_4\otimes I$,
$\tilde{W}_3\otimes T_{34}$
and $I_W\otimes (S_4\oplus P_3)\otimes (I\oplus T_{34})$;
the isovector component as $W_3$
or as the vectors in $\tilde{W}_3\otimes I$, 
$\tilde{V}_4\otimes T_{34}$
and $I_W\otimes (S_4\oplus P_3)\otimes (I\oplus T_{34})$.
The operator with LEC
\begin{eqnarray}
\bar{D}_1^{(1)'}&=&
  \Or\left(e{\tilde \delta}\frac{\mpi^2}{M_{\slashT}^2} \MQ^{-1},
           ew \frac{\mpi^2}{M_{\slashT}^2}\MQ^{-1}\right)
\label{photonscalingC1'}
\end{eqnarray}
transforms as the tensors in 
$\tilde{V}_4\otimes T_{34}$ 
and $I_W\otimes S_4\otimes (I\oplus T_{34})$.
The operators with LECs
\begin{eqnarray}
{\bar S}'^{(1)}_{i}=\Or\left(\frac{ew}{ M_{\slashT}^2 }\MQ^{-1} \right)
\end{eqnarray}
come from $I_W\otimes (I\oplus T_{34})$.
The last operator written explicitly
in Eq. (\ref{eq:cedmEMI}) is a relativistic correction.
It is important to realize that the form of such corrections
depends on the choice of operators included in the effective Lagrangian;
here we have eliminated time derivatives of the nucleon field
through field redefinitions.

For the qEDM, we need also
\begin{eqnarray}
{\cal L}_{\slashT, \gamma N}^{(2)}&=&
\frac{i}{m_N}
\Nb \left(\bar{D}_0^{(1)}+\bar{D}_1^{(1)}\tau_3\right)
S^\mu \mathcal D^\nu_{\perp -} N \, \Fmu 
+\ldots 
\label{qEDMphot2}
\end{eqnarray}
and
\begin{eqnarray}
{\cal L}_{\slashT, \gamma N}^{(3)}&=&
2\Nb \left(\bar{D}_0^{(3)}+\bar{D}_1^{(3)}\tau_3\right) S^\mu N
\, v^\nu \Fmu 
\nonumber\\
&& 
- \bar N \left({\bar S}'^{(3)}_{0}+{\bar S}'^{(3)}_{1}\tau_3\right)
\left(S \cdot \mathcal D_{\perp +} \mathcal D_{\perp +}^{\mu}
      +S^\mu \mathcal D_{\perp +}^2 \right)N 
\, v^\nu \Fmu 
\nonumber\\
&& 
-\frac{1}{4 m_N^2}  
\Nb  \left(\bar{D}_0^{(1)}+\bar{D}_1^{(1)}\tau_3\right)
S \cdot \mathcal D_{\perp -}
\mathcal D_{\perp -}^{\mu} N \, v^{\nu} \Fmu 
+\ldots\ ,
\label{eq:cedmEMII}
\end{eqnarray}
with
\begin{eqnarray}
\bar{D}_i^{(3)}=
\Or\left(e\delta\frac{\mpi^4}{M_{\slashT}^2}\MQ^{-3}\right),
%\\
\qquad
{\bar S}'^{(3)}_{i}= \Or\left(e\delta\frac{\mpi^2}{M_{\slashT}^2}
                                \MQ^{-3}\right).
\label{photonscalingC2}
\end{eqnarray}

With these interactions we can calculate the nucleon EDFF to the order
at which momentum dependence first appears.
We consider a nucleon of initial (final) momentum $p$ ($p'$)
and a (space-like) photon of momentum $q = p-p'$ ($q^2=-Q^2<0$). 
It is convenient to take $q$ and $K=(p+p')/2$ as the independent momenta.
The isoscalar ($F_0$) and isovector ($F_1$)
EDFFs are defined from the nucleon electromagnetic current
$J^\mu_{\mathrm{em}}(q)$
via
\begin{eqnarray}
J^\mu_{\mathrm{em}}(q,K)&=&2\left(F_{0}(Q^2)
+ F_{1}(Q^2)\tau_3\right)
\left\{S^\mu v\cdot q-S\cdot q v^\mu
\right.
\hspace{5.1cm}
\nonumber
\\
&&\left.
+\frac{1}{m_N}\left[q\cdot K S^\mu -S\cdot q K^\mu\right]
+\frac{1}{2m_N^2}S\cdot K\left[K^{\mu} v \cdot q-K \cdot q v^{\mu}\right]
+\ldots
\right\}.
\label{current}
\end{eqnarray}
The first term corresponds to the definition in Ref. \cite{BiraHockings},
while the second is a recoil correction \cite{BiraEmanuele}
and the remaining are consequences of Lorentz invariance.
We will write
\begin{equation}
F_i(Q^2) = D_i -S'_i Q^2+ H_i(Q^2),
\label{eq:Gdef}
\end{equation}
where $D_i$ is the isospin $i$ component of the EDM,
$S'_i$ of the SM, and $H_i(Q^2)$ accounts for the remaining 
$Q^2$ dependence.
The EDFF of the proton (neutron) is $F_0 + F_1$ ($F_0 - F_1$).

\begin{figure}[t]
\centering
\includegraphics[scale = 0.7]{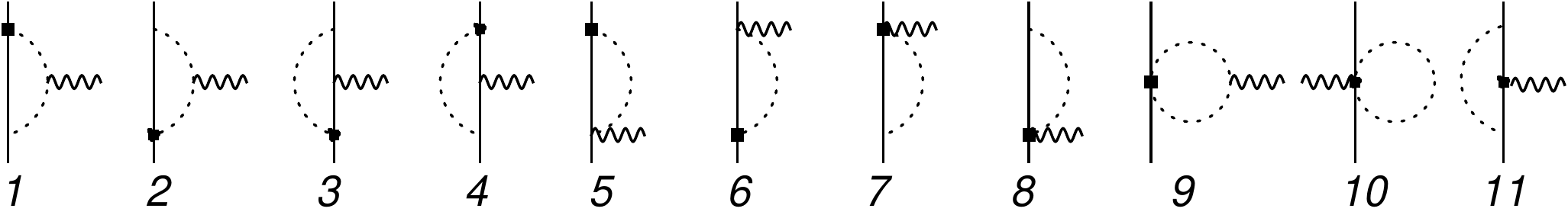}
%{diagramsnucleonedff2.pdf}
\caption{One-loop diagrams contributing to the nucleon EDFF.
Solid, dashed and wavy lines represent the propagation of
nucleons, pions and photons, respectively.
A square marks a $T$-violating interaction,
other vertices representing $T$-conserving interactions.
}\label{EDFFloop}
\end{figure}

The calculation of the EDFF to the order we are interested in
includes $T$ violation in tree and one-loop diagrams.
In tree diagrams the photon is attached to the nucleon line
via a $T$-violating operator from Eqs. 
(\ref{Wphot}), (\ref{Wphot2}), (\ref{eq:cedmEMI}), (\ref{qEDMphot2}), 
and (\ref{eq:cedmEMII}).
The loop diagrams, shown in Fig. \ref{EDFFloop}, contain the $T$-violating 
pion-nucleon couplings in Eq. \eqref{eq:cedm1} or the 
photon-nucleon couplings in Eqs. (\ref{Wphot}) and (\ref{eq:cedmEMI})
---which we denote by squares---  
while the other 
couplings come from the leading,
$T$-preserving chiral Lagrangian, Eq. \eqref{LagrCons}.
In addition, nucleon wave-function renormalization 
from ${\cal L}^{(0)}$ at one-loop level \cite{Bernard:1995dp} can appear.
Since in this Lagrangian the nucleon is static,
in one-loop diagrams we take $v\cdot q = v\cdot K= 0$.
We use dimensional regularization in $d$ dimensions
and encode divergences in the factor
\begin{eqnarray}
L&\equiv& \frac{2}{4-d}-\g_E+\ln 4\pi \ .
\end{eqnarray}
The loops bring in also a renormalization scale $\mu$, which 
is eliminated through the accompanying LECs.

%%%%%%%%%%%%%%%%%%%%%
%%%%%%%%%%%%%%%%%%%%%%

We start with the contributions from the qCEDM, which are very similar
to those of $\bar\theta$ \cite{BiraHockings,Ottnad}.
In this case the lowest-order momentum dependence arises 
from the loops 1-6 in Fig. \ref{EDFFloop},
where the $T$-violating vertex is the first term in Eq. (\ref{eq:cedm1}).
At the same order,
${\cal O}(e\tilde{\delta}m_\pi^2/M_{\slashT}^2\MQ)$,
there are also tree contributions
from the first term in Eq. (\ref{eq:cedmEMI}).
The isoscalar form factor does not receive loop corrections and can be 
expressed purely in terms of
coefficients of short-distance operators,
\begin{eqnarray}
D_{0,\textrm{qCEDM}} &=& {\bar D}_{0}^{(1)}, 
\\
S_{0,\mathrm{qCEDM}}'&=& 0,
\label{Schiffchromo0}\\
H_{0,\textrm{qCEDM}}(Q^2)&=&0.
\label{scalarqCEDM}
\end{eqnarray}
Contributions 
to the isoscalar SM appear in higher orders. 
In contrast, 
the loop diagrams with static nucleons not only
renormalize the contributions of short-distance operators 
to the isovector EDM, but also
generate a non-trivial momentum dependence in the isovector EDFF.
The $\mu$-independent isovector EDM is found to be
\begin{equation}
D_{1, \textrm{qCEDM}} = \bar{D}_1^{(1)}+\bar{D}_1^{(1)'}
+\frac{eg_A\bar{g}_0}{(2\pi\Fp)^2}\left(L-\ln\frac{\mpi^2}{\mu^2} \right),
\end{equation}
while the momentum dependence 
is encoded in  
\begin{eqnarray}
S_{1, \mathrm {qCEDM}}'&=& \frac{eg_A\bar{g}_0}{6\mpi^2(2\pi\Fp)^2},
\label{Schiffchromo1}\\
H_{1,\textrm{qCEDM}}(Q^2)&=& \frac{4eg_A\bar{g}_0}{15(2\pi\Fp)^2} \; 
f\left(\frac{Q^2}{4\mpi^2}\right),
\label{momdepvectorqCEDM}
\end{eqnarray}
where the function $f(Q^2/4 m^2_{\pi})$ is defined as
\begin{equation}\label{eq:function}
f(x)\equiv  
-\frac{15}{4}\left[\sqrt{1+\frac{1}{x}}
\ln \left(\frac{\sqrt{1+\frac{1}{x}}+1}{\sqrt{1+\frac{1}{x}}-1}\right) 
- 2\left(1 +\frac{x}{3}\right)\right].
\end{equation}
Note that $f(x\ll 1)=x^2+{\cal O}(x^3)$. 

Contrary to the qCEDM, the momentum dependence for qEDM and gCEDM 
arises only two orders down with respect to the lowest-order contribution
to the EDM.  To this order, a calculation of the electromagnetic current
yields, in addition to strong-interaction corrections,
also the Lorentz terms $\propto m_N^{-1}$ and
$\propto m_N^{-2}$ in Eq. (\ref{current}). 
In the strong-interaction corrections given below we include
the nucleon wave-function renormalization.

For the qEDM short-range contributions to the EDM 
start at chiral index $\Delta=1$ and others appear at $\Delta =3$.
To this order 
there are no contributions from pion-nucleon $T$-violating interactions, while
the loop diagrams 7, 8, 10, and 11 in Fig.~\ref{EDFFloop},
with interactions from Eq. (\ref{eq:cedmEMI}), 
only renormalize
the tree-level contributions without any energy dependence.  
To ${\cal O}(e\delta m_\pi^4/M_{\slashT}^2\MQ^3)$, we find the EDMs
\begin{eqnarray}
D_{0, \textrm{qEDM}} &=& {\bar D}_{0}^{(1)} +{\bar D}_{0}^{(3)} 
+\frac{3}{4}\bar{D}_0^{(1)}\frac{\mpi^2}{(2\pi\Fp)^2}
 \left[(2+ 4 g_A^2)\left(L-\ln \frac{\mpi^2}{\mu^2}\right)+2 + g_A^2\right],
\label{D0qEDM}
\\
 D_{1, \textrm{qEDM}} &=& \bar{D}_1^{(1)} +\bar{D}_1^{(3)}
+\frac{1}{4}\bar{D}_1^{(1)}\frac{\mpi^2}{ (2\pi\Fp)^2}
 \left[(2 + 8 g_A^2)\left(L-\ln \frac{\mpi^2}{\mu^2}\right)+2 + 3 g_A^2\right],\hspace{5mm}
\label{D1qEDM}
\end{eqnarray}
and the momentum dependence given entirely by the SMs,
\begin{eqnarray}
S'_{i, \mathrm{qEDM}} =  {\bar S}'^{(3)}_{i},
\\
H_{i, \textrm{qEDM}}(Q^2)= 0.
\label{momdepqEDM}
\end{eqnarray}

In the case of the gCEDM, short-range contributions to the EDFF start at 
$\Delta=-1$, which dominate, and appear again at $\Delta =1$,
suppressed by  $m_\pi^2/\MQ^2$.
At this order there are also contributions from the $T$-violating
pion-nucleon interactions in Eq. (\ref{eq:cedm1}) through the loops
1-10 in Fig. \ref{EDFFloop},
and from the photon-nucleon interactions in Eq. (\ref{Wphot})
through the loops 10 and 11.
Thus, to ${\cal O}(e w m_\pi^2/M_{\slashT}^2\MQ)$ we find
the $\mu$-independent EDMs
\begin{eqnarray}
D_{0, \rm{gCEDM}} &=& {\bar D}_{0}^{(-1)} + {\bar D}_{0}^{(1)} 
+  3 g_A^2 {\bar D}_{0}^{(-1)}
\frac{\mpi^2 }{(2\pi\Fp)^2}
\left(L-\ln \frac{\mpi^2}{\mu^2}\right)\ ,
\\
D_{1, \rm{gCEDM}} &=& \bar{D}_1^{(-1)} + \bar{D}_1^{(1)} + \bar{D}_1^{(1)'}
+\frac{m^2_{\pi} }{(2\pi F_{\pi})^2}  
\left\{\left(1+g_A^2\right)\bar{D}_1^{(-1)}+\frac{e\bar \imath_0}{8}
\right.
\nonumber\\
 &&
\left.+ \left[\left(1+ 2g_A^2\right) \bar{D}_1^{(-1)}
 +e\left(\frac{\bar{g}_0g_A}{m^2_{\pi}} +\frac{\bar \imath_0}{8}\right)\right]
\left(L - \ln \frac{m^2_{\pi}}{\mu^2}\right)
\right\}.
\end{eqnarray}
The isoscalar momentum dependence is entirely due to short-range operators
in Eq. \eqref{eq:cedmEMI}, 
\begin{eqnarray}
S'_{0, \textrm{gCEDM}}&=& {\bar S}'^{(1)}_{0}.
\label{WSchiff0}
\\
H_{0, \textrm{gCEDM}}(Q^2)&=& 0.
\label{momdepscalarqCEDM}
\end{eqnarray}
The isovector part, on the other hand,
receives also non-analytic contributions:
\begin{eqnarray}
S'_{1, \textrm{gCEDM}}&=& 
{\bar S}'^{(1)}_{1}
     + \frac{e}{6(2\pi F_{\pi})^2} 
     \left[-\frac{\bar \imath_0}{8}\left(L - \ln \frac{m^2_{\pi}}{\mu^2}\right)
+ \frac{g_A \bar{g}_0}{\mpi^2}\right],
\label{WSchiff1}
\\
H_{1, \textrm{gCEDM}}(Q^2)&=& 
\frac{4e\mpi^2}{15(2\pi F_{\pi})^2}
\left\{\left(\frac{g_A\bar{g}_0}{m_\pi^2}+\frac{\bar \imath_0}{12}\right)
       \; f\left( \frac{Q^2}{4 m^2_{\pi}} \right)
\right.
\nonumber \\
&& \left. 
+\frac{\bar \imath_0}{12}\frac{Q^2}{4m_\pi^2} 
 \left[-\frac{5}{2}\frac{Q^2}{4m_\pi^2}
+f\left(\frac{Q^2}{4 m^2_{\pi}}\right)\right]\right\},
\label{momdepvectorgCEDM}
\end{eqnarray}

We are now in position to discuss the implications of the various
dimension-6 $T$-violation sources to the nucleon EDFF. 

First, we note that to this order the nucleon EDFF stemming from the qCEDM has
a form that is identical to that  
\cite{BiraHockings, Ottnad} from the $\tb$ term.
In both cases the momentum dependence (and thus the SM) is
isovector, has a scale (relative to the EDM) set by $2m_\pi$,
and is determined by the lowest-order pion 
nucleon coupling $\bar{g}_0$.
The EDFF depends on just three independent combinations of LECs, 
$\bar{g}_0$ and the short-range 
EDM contributions ${\bar D}_{0}^{(1)} $ and 
${\bar D}_{1}^{(1)}+\bar{D}_1^{(1)'}$,
which contain nucleon matrix elements of $\tilde{V}_4$ for qCEDM 
and $P_4$ for the $\tb$ term. The numerical factors relating these couplings to
either $\tilde{\delta}$ or $\tb$ will thus be different.
In the case of $\tb$, the matrix element in $\bar{g}_0$ can be determined from
$T$-conserving observables, because it
is related \cite{BiraEmanuele} to the matrix element of $P_3$ that generates
the quark-mass contribution to the nucleon mass splitting:
$\bar{g}_0/\tb \simeq 3$ MeV.
For the qCEDM, an argument identical to that in Ref. \cite{CDVW79}
serves to estimate $D_{1, \rm qCEDM}$ in terms of $\bar{g}_0$,
but no analogous constraint exists for $\bar{g}_0$ in this case
and without a lattice calculation or a model
we cannot do better than dimensional analysis.
(For an estimate with QCD sum rules, see Ref. \cite{khriplo}.)
In any case, to the order we consider here,
any EDFF measurement alone will be equally well reproduced
by a certain value of $\tb$ or a certain value of $\tilde{\delta}$.
Note that the qCEDM does give rise to additional effective interactions
generated by $\tilde{W}_3$, which contribute to the nucleon EDFF 
only at higher orders
but could generate sizable differences for other observables.

Second, the pion-nucleon sector of the qEDM is suppressed compared to that of 
the qCEDM because of the
smallness of $\alpha_{\rm em}$ compared to $g_s^2/4\pi$
at low energies. The consequence is that, up to the lowest order
where momentum dependence appears, both the EDM and the SM 
from the qEDM are determined
by four combinations of six independent LECs, which at this point
can only be estimated by dimensional analysis.
The momentum dependence is expected to be governed by the QCD scale 
$\MQ$,
small relative to the EDM, and nearly linear in $Q^2$.

Finally, in the case of the gCEDM loops are also suppressed,
but do bring in non-analytic terms not only 
to isoscalar and isovector EDMs, but also to the isovector momentum dependence
(and thus SM). 
Again the momentum dependence is governed by $\MQ$.
In addition to seven short-range contributions
to the EDMs and SMs, also 
%three 
two independent pion-nucleon LECs appear 
($\bar{g}_0$ and $\bar{\imath}_0$) which endow the isovector EDFF with
a richer momentum dependence than in other cases.
The isoscalar momentum dependence is identical to qEDM.
For the gCEDM, using the pion loop  
together with an estimate of $\bar{g}_0$ \cite{gudkov}
is likely to be an underestimate of the EDM, because chiral symmetry
allows a short-range 
contribution that is larger by a factor $M_{QCD}^2/m_\pi^2$.

As it is clear from Eqs. (\ref{scalarqCEDM}), (\ref{momdepvectorqCEDM}), 
(\ref{momdepqEDM}), (\ref{momdepscalarqCEDM}), and (\ref{momdepvectorgCEDM}),
the full EDFF momentum dependences 
(for example, the second derivatives of $F_i$ with respect to
$Q^2$) are different for qCEDM (and $\tb$), qEDM, and gCEDM.
Although the isoscalar components all have linear dependences in $Q^2$
(with different slopes)
to the order considered here,
the isovector components show an increasingly complex structure
as one goes from qEDM to $\tb$ and qCEDM to gCEDM.
Determination of nucleon EDMs and SMs alone would not be enough to separate
the four sources, yet they would yield clues.
Expectations about the orders of magnitude 
of various dimensionless quantities are summarized in Table \ref{table1}.

\begin{table}
\caption{Expected orders of magnitude for 
the neutron EDM (in units of $e/\MQ$), the ratio of proton-to-neutron EDMs,
and the ratios of the proton and isoscalar SMs (in units of 
$1/m_\pi^2$) to the neutron EDM, for the $\tb$ term 
\cite{BiraHockings, Ottnad} and for the three dimension-6 sources 
of $T$ violation discussed in the text.}
\begin{tabular}{||c|cccc||} 
\hline\hline
Source & $\tb$ & qCEDM & qEDM & gCEDM \\ 
\hline
$\MQ \, d_n/e$  & 
${\cal O}\left(\tb \frac{\mpi^2}{\MQ^2}\right)$ & 
${\cal O}\left({\tilde \delta} \frac{\mpi^2}{M_{\slashT}^2}\right)$ &
${\cal O}\left(\delta \frac{\mpi^2}{M_{\slashT}^2}\right)$ & 
${\cal O}\left(w \frac{\MQ^2}{M_{\slashT}^2}\right)$ \\
$d_p/d_n$ & 
${\cal O}\left(1\right)$ & 
${\cal O}\left(1\right)$ & 
${\cal O}\left(1\right)$ & 
${\cal O}\left(1\right)$ \\
$\mpi^2 S'_p/d_n$ & 
${\cal O}\left(1\right)$ & 
${\cal O}\left(1\right)$ &
${\cal O}\left(\frac{\mpi^2}{\MQ^2}\right)$ & 
${\cal O}\left(\frac{\mpi^2}{\MQ^2}\right)$ \\ 
$\mpi^2 S'_0/d_n$ & 
${\cal O}\left(\frac{m_\pi}{\MQ}\right)$ & 
${\cal O}\left(\frac{m_\pi}{\MQ}\right)$ & 
${\cal O}\left(\frac{\mpi^2}{\MQ^2}\right)$ & 
${\cal O}\left(\frac{\mpi^2}{\MQ^2}\right)$ \\
\hline\hline
\end{tabular}
\label{table1}
\end{table}

In the first line of Table \ref{table1} one finds the expected
NDA size of the neutron EDM. As it is well known \cite{Pospelov:2005pr},
this is consistent with many other estimates,
such as $d_n=\Or(d_i)$ in the constituent quark model,
and $d_n=\Or(e \tilde{d}_i/4\pi, e d_W \MQ/4\pi)$
from QCD sum rules. 
If $\tilde \delta \sim \delta \sim w ={\cal O}(1)$
(as would be the case for $g_s\sim 4\pi$ and no small phases), then the gCEDM
gives the biggest dimension-6 contribution to the EDFF
because of the chiral-symmetry-breaking suppression
$\Or(m_\pi^2/\MQ^2)$ for the qCEDM and qEDM.
However,
models exist (for example, Ref. \cite{arnowitt})
where $\delta$ and $\tilde \delta$ are enhanced relative to $w$,
and all three sources produce EDFF contributions of the same overall magnitude.
Even so, there is no {\it a priori}
reason to expect cancellations among the various sources.
A measurement of the neutron EDM $d_n$ could be fitted by any one source.
Conversely, barring unlikely cancellations,
the current bound yields order-of-magnitude bounds
on the various parameters at
the scale where NDA applies: using $2\pi F_\pi\simeq 1.2$ GeV for $\MQ$, 
\begin{eqnarray}
\tb&\simle& 10^{-10},
\label{thetabound}\\
\frac{ {\tilde \delta}}{M_{\slashT}^2},\frac{\delta}{M_{\slashT}^2} 
&\simle& \left(10^{5} \; {\rm GeV}\right)^{-2},
\\
\frac{ w}{M_{\slashT}^2} &\simle& \left(10^{6} \; {\rm GeV}\right)^{-2}.
\end{eqnarray}
(For comparison, Eq. (\ref{thetabound}) is consistent within a factor of a few
with 
bounds obtained by taking representative values of $\mu$ in the non-analytic 
terms to estimate \cite{CDVW79}
the size of the renormalized LECs for the EDM, 
and using either $SU(2)$ \cite{BiraEmanuele}
or $SU(3)$ \cite{Ottnad} symmetry to constrain $\bar{g}_0$.) 
In all four cases we expect the proton and neutron
EDMs to be comparable, $|d_p|\sim |d_n|$,
but the presence of undetermined LECs does not allow
further model-independent statements.

It is in the pattern of the $S'_i$ that we see some texture.
(This pattern is not evident in Ref. \cite{otherFaessler},
possibly because of the way chiral symmetry is broken 
explicitly in the model used, both 
in the form of the $T$-conserving pion-nucleon Lagrangian
and in the gCEDM magnitude of the $T$-violating pion-nucleon coupling.)
While in all cases one expects $|S'_p|\sim |S'_n|$,
the relative size to the EDMs, in particular 
of the isovector component, allows one in principle to 
separate qEDM and gCEDM from $\tb$ and qCEDM.
Since all these sources generate different pion-nucleon interactions
thanks to their different chiral-symmetry-breaking properties,
nuclear EDMs might provide further probes of the hadronic source
of $T$ violation.

More could be said with input from lattice QCD.
For each source the pion-mass dependence is different.
A fit to lattice data on the $Q^2$ {\it and} $m_\pi^2$
dependences of the nucleon EDFF with the expressions of this paper
would allow in principle
the separate determination of LECs.
In this case a measurement of the neutron and proton alone
would suffice to pinpoint a dominant source if it exists,
but in the more general case of two or more comparable sources
further observables are needed.

One should keep in mind that our approach is limited to low energies.
The contributions associated with quarks heavier than up and down
are buried in the LECs,
as done, for example, in other calculations of nucleon form factors:
electric and magnetic \cite{lewis}, anapole \cite{anapole},
and electric dipole from $\bar\theta$ \cite{BiraHockings}.
Heavy-quark EDMs and CEDMs are also singlets under 
$SU(2)_L\times SU(2)_R$, so they generate in two-flavor ChPT interactions
with the same structure as those from the gCEDM,
and cannot be separated explicitly from the latter.
(This is clear already in the one-loop running of $d_W$,
which gets a contribution of the heavy-quark CEDMs \cite{Weinberg:1989dx}.)
The parameter $w$ here 
%Our $d_W$ 
should be interpreted as subsuming heavier-quark EDMs and CEDMs.
With the additional assumption that 
$m_s$ makes a good expansion parameter, 
effects of the $s$ quark could be included explicitly.
The larger
$SU(3)_L\times SU(3)_R$ symmetry would yield further relations
among observables 
(for example, between the EDFFs of the nucleon and of the $\Lambda$),
and we could, in principle, isolate
the contributions of the strange quark.
Since our nucleon results, which can be used as input in nuclear calculations
in two-flavor nuclear EFT,
would be recovered in the low-energy limit
anyway ---as was explicitly verified in Ref. \cite{Ottnad}
for the $\bar\theta$ results of Ref. \cite{BiraHockings}---
we leave a study of 
the identification of explicit $s$-quark effects 
to future work.

In summary, we have investigated the low-energy electric
dipole form factor
that emerges as a consequence of effectively dimension-6 sources
of $T$ violation at the quark-gluon level: the quark electric 
and color-electric dipole moments,
and the gluon color-electric dipole moment.
Only the full momentum dependence could in principle separate these sources,
although the Schiff moments, if they were isolated, 
would partially exhibit a texture of $T$ violation.
Further implications of the different chiral-symmetry-breaking patterns
of these sources will be studied in a forthcoming paper \cite{morejordy}.

\acknowledgments
We thank D. Boer, W. Hockings, X. Ji, and M. Ramsey-Musolf for useful discussions.
UvK thanks the hospitality extended to him at KVI, University of Groningen,
and at the Institute for Nuclear Theory, University of Washington.
This research was supported 
by the Dutch Stichting voor Fundamenteel
Onderzoek der Materie (FOM) under programmes 104 and 114 (JdV, RGET)
and by the US DOE under grants
DE-FG02-06ER41449 (EM) and DE-FG02-04ER41338 (EM, UvK).


\begin{thebibliography}{99}

\bibitem{KhripLam1997}
I. B. Khriplovich and S. K. Lamoreaux,
\textit{CP Violation Without Strangeness: Electric Dipole Moments
of Particles, Atoms, and Molecules\/} (Springer Verlag, Berlin, 1997).

\bibitem{Pospelov:2005pr}
M. Pospelov and A. Ritz,
Ann. Phys.\ {\bf 318}, 119 (2005).

\bibitem{Kobayashi:1973fv}
M.~Kobayashi and T.~Maskawa,
Prog.\ Theor.\ Phys.\  {\bf 49}, 652 (1973).

\bibitem{expts}
T. M. Ito, 
J. Phys. Conf. Ser. {\bf 69}, 012037 (2007), nucl-ex/0702024;
K. Bodek {\it et al.}, 
arXiv:0806.4837.

\bibitem{dnbound}
C. A. Baker {\it et al.}, 
Phys. Rev. Lett. {\bf 97}, 131801 (2006).

\bibitem{storageringexpts}
F. J. M. Farley {\it et al.},
Phys. Rev. Lett. {\bf 93}, 052001 (2004);
Y. F. Orlov, W. M. Morse, and Y. K. Semertzidis,
Phys. Rev. Lett. {\bf 96}, 214802 (2006).
%Y. K. Semertzidis, 
%Lect. Notes Phys. {\bf 741}, 97 (2008).

\bibitem{hgbound}
W. C. Griffiths {\it et al.}, 
Phys. Rev. Lett. {\bf 102}, 101601 (2009).

\bibitem{McKellar:1987tf}
I. B. Khriplovich and A. R. Zhitnitsky,
Phys.\ Lett.\  B {\bf 109}, 490 (1982);
B. H. J. McKellar, S. R. Choudhury, X.-G. He, and S. Pakvasa,
Phys.\ Lett.\  B {\bf 197}, 556 (1987);
X.-G. He, B. H.J. McKellar, and S. Pakvasa,
Int. J. Mod. Phys. {\bf A4}, 5011 (1989);
{\bf A6}, 1063(E) (1991).

\bibitem{Pospelov:1994uf}
M. E. Pospelov,
Phys.\ Lett.\  B {\bf 328}, 441 (1994);
A. Czarnecki and B. Krause,
Phys.\ Rev.\ Lett.\ {\bf 78}, 4339 (1997).

\bibitem{BiraHockings} 
W. H. Hockings and U. van Kolck, 
Phys. Lett. B {\bf 605}, 273 (2005).

\bibitem{Faessler}
J. Kuckei, C. Dib, A. F\"a{\ss}ler, T. Gutsche, S. Kovalenko,
V. E. Lyubovitskij, and K. Pumsa-ard,
Phys. Atom. Nucl. {\bf 70}, 349 (2007).

\bibitem{Ottnad}
K. Ottnad, B. Kubis, U.-G. Mei{\ss}ner, and F.-K. Guo,
arXiv:0911.3981.

\bibitem{lattice}
F. Berruto, T. Blum, K. Orginos, and A. Soni,
Phys. Rev. D {\bf 73}, 054509 (2006);
E. Shintani, S. Aoki, and Y. Kuramashi, 
Phys. Rev. D {\bf 78}, 014503 (2008);
R. Horsley {\it et al.},
arXiv:0808.1428.

\bibitem{weinberg79} 
S. Weinberg, 
Physica {\bf 96A}, 327 (1979);
J. Gasser and H. Leutwyler, 
Ann. Phys. {\bf 158}, 142 (1984);
Nucl. Phys. {\bf B250}, 465 (1985).

\bibitem{original}
S. Weinberg,
Phys. Lett. B {\bf 251}, 288 (1990);
Nucl. Phys. B {\bf 363}, 3 (1991).

\bibitem{Jenkins:1990jv}
E. E. Jenkins and A. V. Manohar,
Phys.\ Lett.\  B {\bf 255}, 558 (1991).
  
\bibitem{Weinberg} 
S. Weinberg, 
\textit{The Quantum Theory of Fields\/}, Vol. 2
(Cambridge University Press, Cambridge, 1996).

\bibitem{Bernard:1995dp}
V.~Bernard, N.~Kaiser, and U.~G.~Mei{\ss}ner,
Int.\ J.\ Mod.\ Phys.\  E {\bf 4}, 193 (1995).

\bibitem{BiraEmanuele} 
E. Mereghetti, W. H. Hockings, and U. van Kolck,
Ann. Phys. (to appear), arXiv:1002.2391.

\bibitem{CDVW79}
R. J. Crewther, P. Di Vecchia, G. Veneziano, and E. Witten,
Phys. Lett. B {\bf 88}, 123 (1979); 
{\bf 91}, 487(E) (1980).

\bibitem{su3}
H.-Y. Cheng, 
Phys. Rev. D {\bf 44}, 166 (1991);
A. Pich and E. de Rafael, 
Nucl. Phys. B {\bf 367}, 313 (1991);
P. Cho, 
Phys. Rev. D {\bf 48}, 3304 (1993);
B. Borasoy, 
Phys. Rev. D {\bf 61}, 114017 (2000);
S. Narison, 
Phys. Lett. B {\bf 666}, 455 (2008).

\bibitem{Thomas:1994wi}
S. D. Thomas,
Phys.\ Rev.\  D {\bf 51}, 3955 (1995).

\bibitem{oconnell}
D. O'Connell and M.J. Savage,
Phys. Lett. B {\bf 633}, 319 (2006).

\bibitem{Rujula} 
W. Buchm\"uller and D. Wyler,
Nucl. Phys. B {\bf 268}, 621 (1986);
A. De R\'ujula, M. B. Gavela, O. P\`ene, and F. J. Vegas, 
Nucl. Phys. B {\bf 357}, 311 (1991).

\bibitem{Weinberg:1989dx}
S.~Weinberg,
Phys.\ Rev.\ Lett.\  {\bf 63}, 2333 (1989);
E. Braaten, C. S. Li, and T. C. Yuan,
Phys.\ Rev.\ Lett.\  {\bf 64}, 1709 (1990).

\bibitem{otherFaessler}
C. Dib, A. F\"a{\ss}ler, T. Gutsche, S. Kovalenko, J. Kuckei, 
V. E. Lyubovitskij, and K. Pumsa-ard,
J. Phys. G {\bf 32}, 547 (2006).

\bibitem{will}
W. H. Hockings, Ph.D. dissertation, University of Arizona (2006).

\bibitem{morejordy}
J. de Vries, E. Mereghetti, R. G. E. Timmermans, and U. van Kolck,
in preparation.

\bibitem{Jarlskog}
C. Jarlskog, 
Phys. Rev. Lett. {\bf 55}, 1039 (1985).

\bibitem{arnowitt}
R. Arnowitt, J. L. Lopez, and D. V. Nanopoulos, 
Phys. Rev. D {\bf 42}, 2423 (1990);
R. Arnowitt, M. J. Duff, and K. S. Stelle, 
Phys. Rev. D {\bf 43}, 3085 (1991).

\bibitem{ibrahim}
T. Ibrahim and P. Nath,
Phys.\ Lett.\ B {\bf 418}, 98 (1998);
Phys.\ Rev.\ D {\bf 57}, 478 (1998);
{\bf 58}, 019901(E) (1998);
{\bf 60}, 119901(E) (1999);
{\bf 60}, 079903(E) (1999).

\bibitem{ji}
H. An, X. Ji, and F. Xu,
JHEP {\bf 1002}, 043 (2010);
F. Xu, H. An, and X. Ji, 
JHEP {\bf 1003}, 088 (2010).

\bibitem{vanKolck}
U. van Kolck, Ph.D. dissertation, University of Texas (1993);
Few Body Syst. Suppl. {\bf 9}, 444 (1995).

\bibitem{Baluni}
V. Baluni, 
Phys. Rev. D {\bf 19}, 2227 (1979).

\bibitem{ManoharLuke} 
M. E. Luke and A. V. Manohar, 
Phys. Lett. B \textbf{286}, 348 (1992).

\bibitem{NDA}
A. V. Manohar and H. Georgi,
Nucl. Phys. B {\bf 234}, 189 (1984).

\bibitem{lewis}
R. Lewis and N. Mobed,
Phys. Rev. D {\bf 59}, 073002 (1999);
B. Kubis and R. Lewis,
Phys. Rev. C {\bf 74}, 015204 (2006).

V. Bernard, N. Kaiser, J. Kambor, and U.-G. Mei{\ss}ner,
Nucl. Phys. B {\bf 388}, 315 (1992);
V. Bernard, H.W. Fearing, T.R. Hemmert, and U.-G. Mei{\ss}ner,
Nucl. Phys. A {\bf 635}, 121 (1998), (E) {\bf 642}, 563 (1998).

\bibitem{anapole}
C. M. Maekawa and U. van Kolck,
Phys. Lett. B {\bf 478}, 73 (2000);
C. M. Maekawa, J. S. Veiga, and U. van Kolck,
Phys. Lett. B {\bf 488}, 167 (2000).

\bibitem{khriplo}
V. M. Khatsymovsky and I. B. Khriplovich,
Phys. Lett. B \textbf{296}, 219 (1992).

\bibitem{gudkov}
V. P. Gudkov,
Z. Phys. A {\bf 343}, 437 (1992).

\end{thebibliography}
\end{document}